%Paper: hep-th/9308011
%From: adhar@theory.tifr.res.in (Avinash Dhar)
%Date: Wed, 4 Aug 93 14:57:23 IST

%%%%%%%%%%            PLEASE USE PHYZZX                 %%%%%%%%%%%%
%%%%%%%%%%                                              %%%%%%%%%%%%

\pubnum{93-34}
\date{}
\titlepage
\voffset=24pt
\title{\bf Two-Dimensional Black Hole and Nonperturbative String
Theory$^\star$}\foot{Based on invited talk delivered at the Spring Workshop on
``String Theory, Gravity and Gauge Theory'' at ICTP, Trieste, April 28-29,
1993.}
\author{\rm Avinash Dhar}
\address{\rm Theoretical Physics Group \break Tata Institute of
Fundamental Research \break Homi Bhabha Road, Bombay 400 005, India \break
e-mail: adhar@tifrvax.bitnet}
\abstract{We discuss the interpertation of the $c=1$ matrix model as
two-dimensional string theory in a dilaton-black hole background.  The
nonperturbative formulation of $c=1$ matrix model in terms of an
integrable model of nonrelativistic fermions enables us to study the
quantum fate of the classical black hole singularity.  We find that the
classical singularity is wiped out by quantum corrections when summed to
all orders.}

\def\del{\partial}
\def\nl{\hfill\break}
\def\ni{\noindent}
\def\ket{\rangle}
\def\bra{\langle}
\def\u{\,{\cal U}}
\def\uu{\,{\widehat {\cal U}}}
\def\Duv{[4(uv -\mu/2) \del_u \del_v + 2 (u\del_u + v\del_v) + 1]}
\def\barDuv{4(u \bar u -\mu/2) \del_u \del_{\bar u} + 2 (u\del_u +
\bar u\del_{\bar u}) + 1}
\def\gst{{g_{\rm str}}}
\def\T{{\widehat T}}
\def\delt{\delta T}
\def\ddelt{\delta {\widehat T}}
\def\pphi{{\widehat \phi}}
\def\etab{{\bar \eta}}

\def\R{{\cal R}}
\def\x{{\bf x}}
\def\t{{\bf t}}

\REF\ONE{S.R. Das, S. Naik and S.R. Wadia, Mod. Phys. Lett. A4 (1989)
1033; A. Dhar, T. Jayaraman, K.S. Narain and S.R. Wadia, Mod. Phys. Lett.
A5 (1990) 863; S.R. Das, A. Dhar and S.R. Wadia, Mod. Phys. Lett. A5(1990)
799.}
\REF\TWO{J. Polchinski, Nucl. Phys. B234 (1989) 123.}
\REF\THREE{T. Banks and J. Lykken, Nucl. Phys. B331 (1990) 173.}
\REF\FOUR{
E. Brezin, C. Itzykson, G. Parisi and J.B. Zuber, Comm. MAth. Phys. 59
(1978) 35;
E. Brezin, V.A. Kazakov and Al.B. Zamolodchikov, Nucl. Phys. B338 (1990)
673;
D. Gross and N. Miljkovich, Nucl. Phys. B238 (1990) 217;
P. Ginsparg and J. Zinn-Justin, Phys. Lett. B240 (1990) 333;
G. Parisi, Europhys. Lett. 11 (1990) 595.}
\REF\FIVE{G. Mandal, A.M. Sengupta and S.R. Wadia, Mod. Phys. Lett. A6
(1991) 1685.}
\REF\SIX{E. Witten, Phys. Rev. D44 (1991) 314.}
\REF\SEVEN{For a review see A. Strominger  and J. Harvey, Chicago preprint
EFI-92-41, hep-th/9209055.}
\REF\EIGHT{A.M. Sengupta and S.R. Wadia, Int. J. Mod. Phys. A6 (1991) 1961.}
\REF\NINE{D.J. Gross and I. Klebanov, Nucl. Phys. B352 (1990) 671.}
\REF\TEN{G. Moore, Nucl. Phys. B368 (1992) 557.}
\REF\ELEVEN{S.R. Das and A. Jevicki, Mod. Phys. Lett. A5 (1990) 1639.}
\REF\TWELVE{J. Polchinski, Nucl. Phys. B346 (1990) 253.}
\REF\THIRTEEN{A. Dhar, G. Mandal and S.R. Wadia, Mod. Phys. Lett. A7 (1992)
3129.}
\REF\FOURTEEN{S.R. Wadia, Tata Institute preprint TIFR/TH/93-17.  To
appear in the Proceedings of 1992 Nishinomiya-Yukawa Memorial Symposium.}
\REF\FIFTEEN{A. Dhar, G. Mandal and S.R. Wadia, Tata Institute preprint
TIFR/TH/93-33.}
\REF\SIXTEEN{S.R. Das, A. Dhar, G. Mandal and S.R. Wadia, Int. J. Mod. Phys.
A7 (1992) 5165.}
\REF\SEVENTEEN{I. Bakas, Phys. Lett. \underbar{B228} (1989) 57 and Comm.
Math. Phys. \underbar{134} (1990) 487; A. Bilal, Phys. Lett.
\underbar{B227} (1989) 406; C. Pope, L. Romans and X. Shen, Phys. Lett.
\underbar{B236} (1990) 173, Nucl. Phys. \underbar{B339} (1990) 91 and
Phys. Lett. \underbar{B242} (1990) 401.}
\REF\EIGHTEEN{J. Avan and A. Jevicki, Phys. Lett. \underbar{B226} (1991)
35 and Phys. Lett. \underbar{B272} (1991) 17; D. Minic, J. Polchinski and
Z. Yang, Nucl. Phys. \underbar{B362} (1991) 125; G. Moore and N. Seiberg,
Int. J. Mod. Phys. \underbar{A7} (1992) 2601.}
\REF\NINTEEN{E. Witten, Nucl. Phys. \underbar{B373} (1992) 187; I.
Klebanov and A.M. Polyakov, Mod. Phys. Lett. \underbar{A6} (1991) 3273.}
\REF\TWENTY{A. Dhar,  G. Mandal and S.R. Wadia, Int. J. Mod. Phys.
\underbar{A8} (1993) 325.}
\REF\TWENTYONE{A. Dhar, G. Mandal and S.R. Wadia, Tata Institute preprint
TIFR/TH/92-40.  To appear in Int. Jour. of Mod. Phys. A.}
\REF\TWENTYTWO{A. Dhar, G. Mandal and S.R. Wadia, Mod. Phys. Lett. A7 (1992)
3703.}
\REF\TWENTYTHREE{A. Dhar, G. Mandal and S.R. Wadia, Tata Institute preprint
TIFR/TH/93-05.  To appear in Mod. Phys. Lett. A.}
\REF\TWENTYFOUR{E. Martinec and S. Shatashvilli, Nucl. Phys.
\underbar{B368} (1992) 338.}
\REF\TWENTYFIVE{S.R. Das, Mod. Phys. Lett. \underbar{A8} (1993) 69.}
\REF\TWENTYSIX{J.G. Russo, Phys. Lett. \underbar{B300} (1993) 336.}
\REF\TWENTYSEVEN{T. Yoneya, Tokyo (Komaba) preprint UT-KOMABA-92-13.}
\REF\TWENTYEIGHT{S. Mukhi and C. Vafa, Tata Institute and Harvard preprint,
TIFR/TH/93-01.}
\REF\TWENTYNINE{A. Jevicki and T. Yoneya, Brown preprint BROWN-HEP-904,
hep-th/9305109.}
\REF\THIRTY{I.S. Gradshteyn and I.M. Ryzhik, {\sl Table of Integrals,
Series and Products}, Academic Press, New York (1965), p. 1039.}
\REF\THIRTYONE{I.S. Gradshteyn and I.M. Ryzhik, Table of Integrals, Series
and Products, Academic Press,  New York (1965), p. 955.}
\REF\THIRTYTWO{D.J. Gross and N. Miljkovich, ref. [4].}
\REF\THIRTYTHREE{G. Mandal, A.M. Sengupta and S.R. Wadia, Mod. Phys. Lett. A6
(1991) 1465.}
\REF\THIRTYFOUR{K. Demeterfi, A. Jevicki and J. Rodrigues, Nucl. Phys. B362
(1991) 173.}

\endpage

\noindent {\bf 1. \underbar{\bf Introduction}}

Two-dimensional string theory can be viewed as one-dimensional matter
(time) coupled to world-sheet gravity [\ONE-\THREE].  The latter has a
non-perturbative formulation in terms of a one-dimensional matrix model.
Since the double-scaled limit of this model can be mapped on to a
completely integrable model of noninteracting nonrelativistic fermions in
one space dimension [\FOUR], we have a unique opportunity of studying some
basic conceptual question of string theory in a nonperturbative setting.
It is rather fortunate that two-dimensional string theory also possesses a
classical black hole solution [\FIVE-\SEVEN].  This presents us with the
opportunity of studying the quantum fate of the classical black hole
singularity.   However, before such a task can be undertaken one must
contend with the puzzle of how the string theory black hole can
emerge in the nonperturbative setting since the matrix model is formulated
in only one dimension (time) and the equivalent fermion theory in a {\bf flat}
two-dimensional space-time [\EIGHT-\TEN].  It is plausible that the
resolution of
this puzzle lies in the fact that (i) the graviton-dilaton system
$(G_{\mu\nu},\Phi)$ of the two-dimensional black hole is equivalent to the
metric $\tilde G_{\mu\nu} = G_{\mu\nu} \exp(-2 \Phi)$, which corresponds to
a space-time that is flat but has a boundary determined by the condition
$\exp(-2 \Phi) \geq 0$, and (ii) the space-time of the perturbative
collective excitation of the fermion theory is a half-plane, the boundary
of which is associated with the classical turning point of the fermions
[\EIGHT-\TWELVE].  It is also important to realize that in the nonperturbative
setting two-dimensional space-time is an {\bf emergent property} and not an a
priori given thing.  There is, therefore, a possibility that two different
metrics may correspond to two different interpretations of the same
physical theory.  In these notes we describe some recent work that
realizes this last mentioned scenario.

The organization of these notes is as follows.  In the next section we
will review some aspects of the string field theory in two dimensions
discussed in ref. [\THIRTEEN].  This string field theory is constructed by an
exact bosonization of the nonrelativistic fermion theory
[\FOURTEEN,\FIFTEEN].  The
basic variable in the bosonized formulation is the operator phase space
density $\uu(p,q,t)$ of the fermions.  In sec. 3 we introduce a new
scalar field, which is a nonlocal transform of $\uu(p,q,t)$ in phase
space, in terms of which the theory has the interpretation of string field
theory in the background of two-dimensional dilaton -- black hole.  In
sec. 4 we derive a nonlinear equation for the new field and also discuss
the connection of this field with the collective field.  In sec. 5 we argue
that the classical black hole singularity is absent in the exact quantum
theory.  In sec. 6 we discuss some novel features of the quantum theory
of the new scalar field.  In sec. 7 we discuss how the Euclidean black
hole emerges from an analytically continued fermion theory that corresponds to
the right side up harmonic oscillator potential.  Sec. 8 is devoted to
some concluding remarks.
\bigskip

\noindent {\bf 2. \underbar{\bf Review of Two-Dimensional String Field
Theory} }

\nobreak
The two-dimensional string field theory discussed in ref. [\THIRTEEN] is
obtained
by an exact bosonization of the nonrelativistic fermion theory to which
the $c=1$ matrix model can be mapped.  The fermion theory is described by
the action
$$
S[\psi] = \int^{+\infty}_{-\infty} dt \int^{+\infty}_{-\infty} dx~\psi^+
(x,t) [i\partial_t - \hat h] \psi(x,t)
\eqn\one
$$
where the single-particle hamiltonian $\hat h$ is given by
$$
\eqalign{&
\hat h = {1\over2} \left(-\partial^2_x + V(x)\right), \cr &
V(x) = -x^2 + {g_3 \over \sqrt{N}} x^3 + \cdots }
\eqn\two
$$
Here $N$ is the total number of fermions,
$$
N = \int^{+\infty}_{-\infty} dx ~\psi^+ (x,t) \psi(x,t).
$$
The continuum (double scaling) limit is obtained by letting $N \rightarrow
\infty$ and the fermi energy $\epsilon_F \rightarrow 0$ while keeping the
renormalized fermi energy $\mu \sim N\epsilon_F$ fixed.  The string
coupling $\gst$ is then given by $\gst \sim {1 \over \mu}$.  This is
also the Planck's constant $\hbar$ for the fermion theory.

The basic object in terms of which bosonization of \one\ ~is achieved is the
operator phase space density of fermions:
$$
\uu(p,q,t) = \int^{+\infty}_{-\infty} dx~\psi^+(q-x/2,t) e^{-ipx}
\psi(q+x/2,t)
\eqn\three
$$
Expectation value $\u(p,q,t)$ of $\uu(p,q,t)$ in any given fermion state
measures the phase space density of fermions in that state.

The key organizing principle on which the bosonization of \one\ ~is based is
the $W_\infty$ algebra and its representation in terms of nonrelativistic
fermion [\THIRTEEN,\SIXTEEN].  In fact \one\ can be interpreted as  the
fermion representative of
a spin in a magnetic field.  The `spin' is the operator $\uu$, the
`rotation' group is $W_\infty$ and the `magnetic field' is the hamiltonian
$\hat h$ given in \two.  This analogy enables one to obtain the bosonised
action using the coherent state method for $W_\infty$ group
[\FOURTEEN,\FIFTEEN].  We will only summarize the result here and refer
the reader to [\THIRTEEN,\FIFTEEN] for details.  The
bosonised action is given by
$$
\eqalign{
S[u] = \int ds~ & dt \int {dp~dq \over 2\pi} \u(p,q,t,s) \left\{\partial_s
\u(p,q,t,s),\partial_t \u(p,q,t,s)\right\}_{MB} \cr &
+ \int dt \int {dp~dq \over 2\pi} h(p,q) \u(p,q,t)}
\eqn\four
$$
where
$$
h(p,q) = {1\over2} (p^2 + V(q)), ~~~V(q) = -q^2 + {g_3 \over \sqrt{N}} q^3
+ \cdots
\eqn\five
$$
In \four\ ~$\u(p,q,t,s)$ is a one-parameter $(s)$ extension of $\u(p,q,t)$
such that for $-\infty < t < \infty$ the range of $s$ is $-\infty < s \leq
0$ with $\u(p,q,t) = \u(p,q,t,s)|_{s=0}$ while $\u(p,q,t,s)|_{s=-\infty}$ is
some time-independent function of $(p,q)$.  Also, the Moyal bracket
$\{~~~\}_{MB}$ is defined as
$$
\left\{A(p,q),B(p,q)\right\}_{MB} = \sin {1\over2} (\partial_p
\partial_{q'} -\partial_q \partial_{p'}) [A(p,q) B(p',q')]\bigg|_{p'=p
\atop q' = q}
\eqn\six
$$
The reader will recognize the 1st term in \four\ as the $W_\infty$ analogue
of the `solid angle' term in the action of an $SU(2)$ spin.  The second
term in \four\ is the analogue of the coupling to magnetic field.

The action \four\ must be supplemented with constraints in the functional
integral measure.  These constraints are
$$
\cos {1\over2} (\partial_p \partial_{q'} - \partial_q \partial_{p'})
[\u(p,q,t) \u(p',q',t)]\bigg|_{p'=p \atop q'=q} = \u(p,q,t)
\eqn\seven
$$
$$
\int {dp~dq \over 2\pi} \u(p,q,t) = N
\eqn\eight
$$

The action \four, together with the constraints \seven\ ~and \eight,
{}~defines
two-dimensional string field theory and can be used to obtain any
correlation function of the $\uu$'s.  Loop expansion is obtained by
letting $p \rightarrow p/\sqrt{\kappa}$, $q \rightarrow q/\sqrt{\kappa}$ and
expanding in powers of $\kappa^2$.  In the classical limit $\gst
\rightarrow 0$, the $W_\infty$ algebra goes over to the algebra $\omega_\infty$
of area-preserving diffeomorphisms in two-dimensions, which is then the
symmetry algebra of the classical theory [\THIRTEEN,\SEVENTEEN -\TWENTY].
The constraint \seven\ then goes over to
$$
\left(\u(p,q,t)\right)^2 = \u(p,q,t)
\eqn\nine
$$
with solutions which are characteristic functions of regions in phase
space.  So, the classical physics is described by fermi fluid `droplets'
in phase space, which move according to the equation
$$
\partial_t \u(p,q,t) + \{h(p,q),\u(p,q,t)\}_{_{PB}} = 0
\eqn\ten
$$
This equation of motion can be derived by verying \four\ preserving the
constraints \seven\ ~and
\eight\ ~and taking the classical limit of the resulting equation.  In the
double scaling limit \ten\ ~becomes
$$
(\partial_t + p\partial_q + q\partial_p) \u(p,q,t) = 0
\eqn\eleven
$$
For the (classical) fermi ground state $\u(p,q,t) = \u_0(p,q)$ is a function
of the classical energy ${1\over2} (p^2-q^2)$.  So the profile of the
characterstic function that describes the fermi ground state is the
hyperbola given by ${1\over2}(p^2-q^2) = \mu$.
Here $\mu$ is the fermi energy (which we will assume to be negative) so
that $\gst \sim {1 \over |\mu|}$.  Small fluctuation
around fermi ground state describe a massless particle.  It is worth
noting that neither
the action \four\ ~nor the constraints \seven\ ~and \eight\ ~`know' about the
string coupling $\gst$.  It makes its first appearence as a parameter
in the solution for fermi ground state and enters the physics of
fluctuations through it.  For further details the reader is referred to
[\THIRTEEN].  The precise nature of the truncation of the full theory that
leads
to collective field theory is discussed in [\TWENTYONE] and reviewed in
[\FOURTEEN,\FIFTEEN].
\bigskip

\noindent {3. \underbar{\bf Hyperbolic Transform}}

\nobreak
In this section we introduce the new fermion bilinear $\pphi(p,q,t)$.
This is defined by a
`hyperbolic transform' of the quantum phase space density $\uu(p,q,t)$
[\TWENTYTWO,\TWENTYTHREE]:
$$ \pphi(p,q,t) = \int dp'dq'\, K(p,q| p',q') \uu(p',q',t)
\eqn\twelve$$
where
$$ K(p,q| p',q') \equiv | (p-p')^2 - (q-q')^2 |^{-1/2}
\eqn\thirteen$$
Equation \twelve\ ~can be regarded as a relation between Heisenberg
operators, where
$$\pphi(p,q,t)= e^{iHt} \pphi(p,q,0) e^{-iHt},\; \uu(p,q,t) = e^{iHt}
\uu(p,q,0) e^{-iHt} \eqn\fourteen$$
The hamiltonian $H$ is given by
$$H = \int {dp dq\over 2\pi}
h(p,q) \uu(p,q,t),\qquad h(p,q) = {1\over 2}(p^2-q^2)
\eqn\fifteen$$
Clearly, equation \twelve\ ~is also valid for expectation values
$$ \bra \psi|\pphi(p,q,t)|\psi \ket = \int dp'dq'\, K(p,q| p',q')
\bra \psi|\uu(p',q',t) | \psi \ket
\eqn\sixteen$$
where $| \psi \ket $ is any state in the Fermi theory.

For the Heisenberg operator $\uu(p,q,t)$ we have the equation of motion
$$ (\del_t + p\del_q + q\del_p)\uu(p,q,t) = 0  \eqn\seventeen $$
Used in the definition \twelve, together with \thirteen, this leads to
$$ (\del_t + p\del_q + q\del_p)\pphi(p,q,t) = 0  \eqn\eighteen $$
The last equation implies that if we define the variables
$$ u = e^{-t} (p +q)/2,\; v= e^t (p-q)/2  \eqn\nineteen $$
or, equivalently,
$$ p = ue^t + ve^{-t},\; q= ue^t - ve^{-t} \eqn\twenty $$
then, at fixed $u$ and $v$,
$$ \del_t \pphi(ue^t + ve^{-t}, ue^t - ve^{-t},t) = 0 \eqn\twentyone$$
This means that we can define the operator
$$\eqalign{
\T(u,v) \equiv & \pphi(ue^t + ve^{-t}, ue^t - ve^{-t},t)  \cr
=&\int du' dv' {\widetilde K}(u,v | u',v')
\uu(u'e^t+v'e^{-t},u'e^t-v'e^{-t},t), \cr}
\eqn\twentytwo $$
which is actually independent of $t$. Here
$${\widetilde K}(u,v| u',v') \equiv\, |(u - u') ( v - v')|^{-1/2}
\eqn\twentythree $$

Before launching into properties of the `tachyon field' \twentytwo\ ~it
is useful to understand its definition in some more detail.  Below
we list some remarkable properties of the kernel
\thirteen ~that define for us the `hyperbolic transform' \twelve.
The basic properties of \thirteen ~(equivalently, of \twentythree ) ~are: \nl
\ni (i) Lorentz covariance: $K(P, Q | P', Q') = K(p,q| p',q')$, where $P =
p\cosh \theta + q\sinh \theta, Q = p\sinh \theta + q \cosh \theta$ and
similarly for $P',Q'$. \nl
\ni (ii) Translation invariance: $K(p,q| p',q') = f(p-p', q-q')$ \nl
\ni (iii) Differential equation:
$$\eqalign{
\,&\Duv \widetilde K(u,v | u',v') = O[u'v'- \mu/2] \cr
\,&\widetilde K(u,v | u',v') = K({u+v\over 2},{u-v\over 2}|{u'+v'\over 2},
{u'-v'\over 2}) \cr}
\eqn\twentyfour$$

The precise form of the last equation is
$$\eqalign{
\Duv & \widetilde K(u,v | u'v') \cr
\,&= 4(u'v'-\mu/2) \del_u\del_v \widetilde K(u,v|u',v')\cr}
\eqn\twentyfive$$
This last property of the kernel is responsible for
the black hole interpretation of the low energy physics of the tachyon.
To see this let us first define the `tachyon fluctuation operator'
$\delta\hat T(u,v)$.  Let $|\psi_0\rangle$ be the ground state and
$|\psi\rangle$ any exicted state of the fermion theory.  Then, we define
$$
\hat T(u,v) - \langle \psi_0|\hat T(u,v)|\psi_0\rangle \equiv \delta\hat
T(u,v),
\eqn\twentysix
$$
and also its expectation value
$$
\langle \psi|\delta\hat T(u,v)|\psi\rangle \equiv \delta T(u,v)
\eqn\twentyseven
$$
In the $\gst \rightarrow 0$ limit the ground state $|\psi_0\rangle$ is
represented by the fermi surface $p^2 - q^2 = 2\mu$.  For states
$|\psi\rangle$ that differ from the ground state $|\psi_0\rangle$ at most
in a small neighbourhood of the
fermi surface one can show, using \twentyfive, that $\delta T(u,v)$
satisfies the equation
$$
[4(uv - \mu/2) \partial_u\partial_v + 2(u\partial_u + v\partial_v) + 1]
\delta T(u,v) \equiv O + O\left({\delta E \over \mu}\right)
\eqn\twentyeight
$$
The $O(\delta E/\mu)$ corrections are small if the fluctuations around the
fermi surface are small.  We shall see the precise nature of the
corrections in the next section.  The leading order equation is precisely
that of tachyon in the dilaton-black hole metric [\FIVE] given by
$$
ds^2 = {du~dv \over (uv - \mu/2)},
$$
the dilaton field $\Phi$ being given by
$$
e^{-2\Phi} = uv - \mu/2.
$$

Clearly, property (iii) is very  desirable from the point of view of black
hole physics.  Property (i) implies that the equations of motion are the
same for $\hat \phi(p,q,t)$ and $\uu(p,q,t)$; this implies that the reduced
variables $u,v$ used in both cases have the same physical interpretation.
Property (ii) is directly related to the fact that the hyperbolic
transform becomes local in Fourier space; in other words, the (double)
Fourier transform of $\hat \phi(p,q,t)$'s ($\hat {\widetilde \phi}(\alpha,
\beta, t)$'s) are basically rescalings of $W(\alpha,\beta,t)$'s [\THIRTEEN]
which ensures that the $\hat{\widetilde \phi}$'s have an
algebra that has
a classical limit. The latter property implies that the classical action
written in terms of $\hat\phi(p,q,t)$ has a $\gst \to 0$
limit which, once again, is rather crucial.

We now prove that \thirteen ~is the {\bf unique kernel} satisfying all the
three properties mentioned above.

\ni {\bf Proof:} \nl
\ni Properties (i) and (ii) imply that $K$ is some function of only the
combination $(p-p')^2 - (q-q')^2$:
$$
K(p,q | p', q') = g((p-p')^2- (q-q')^2)) \longleftrightarrow
\widetilde K(u,v | u',v') =  f(x),\;
x= (u-u')(v-v')
\eqn\twentynine
$$
Property (iii) states that if we choose
$u' = q_0 e^\theta/2, v'= -q_0 e^{-\theta}/2, q_0\equiv
\sqrt{-2\mu}$ so that
$u'v' = \mu/2$ (note that $\mu$, the fermi energy, is negative in our
convention), then
$$\Duv \widetilde K(u,v | q_0e^\theta/2, -q_0e^{-\theta}/2) = 0
\eqn\thirty$$
For this choice of $u',v'$ we have $x = uv -\mu/2 + [q_0/2] (ue^{-\theta}
- ve^\theta)$. Using \twentyeight\ ~for $\widetilde K$, and introducing the
notation $y = uv - \mu/2$ , we get from \twentynine
$$ 2xf'(x) + f(x) + y[ 6f'(x) + 4xf''(x)]= 0
\eqn\thirtyone$$
This equation must be identically satisfied for all $y$, which implies
that both the $y$-independent term and the coefficient  of $y$ must vanish
in \thirty . Curiously the first condition implies the second condition, so
we need not separately consider the second condition. Thus we get
$$ 2xf'(x) + f(x) = 0
\eqn\thirtytwo$$
The above equation is solved by
$$ f(x) = \hbox{constant}\; |x|^{-1/2}
\eqn\thirtythree$$
which proves that (modulo an overall constant)
$$ K(p,q|p',q') = |(p-p')^2 -(q-q')^2|^{-1/2}
\longleftrightarrow \widetilde K(u,v| u',v') = |(u-u')(v-v')|^{-1/2}$$

To end this section we mention other related works
[\TWENTYFOUR-\TWENTYNINE].  Refs. [\TWENTYFOUR] and [\TWENTYEIGHT] deal
with the continuum formulation and relate the coset black holes with
Liouville theory.  Refs. [\TWENTYFIVE-\TWENTYSEVEN] and [\TWENTYNINE] deal
directly with the matrix model.  We also mention here that the criticism
of our work in ref. [\TWENTYNINE] is really not valid.  This is because
with both the cosmological constant and black hole mass perturbations
present there is no simple scaling by which the dependence of the theory
on either of the parameters appears solely through the string coupling, as
is the case when only either one of the perturbations is present.  In
fact, with both the perturbations present there is nothing in the theory
that prevents us from setting black hole mass equal to the cosmological
constant.  We believe that the $c=1$ matrix model describes precisely that
situation.
\bigskip

\noindent {\bf 4. \underbar{\bf Non-linear Differential Equation for the
Tachyon}}

\nobreak
In this section we discuss the semiclassical physics of the tachyon
$\T(u,v)$ in detail and derive a closed
non-linear differential equation
for $\delt(u,v)$  (defined in \twentysix\ ~and \twentyseven ) in a
semiclassical expansion.

Let  us consider states $| \psi \ket $ which satisfy
$$ \bra \psi | \uu(p,q,t) | \psi \ket = \vartheta ([ p_+(q,t)- p] [p -
p_-(q,t)]) + O(\gst).  \eqn\thirtyfour $$
where $\vartheta(x) = 1$ if $x>0$ and $=0$ otherwise.
This corresponds to the `quadratic profile' ansatz [\TWELVE,\TWENTY].
We recall that
in the $g_{str} \to 0$ limit, the classical $\u(p,q,t)$ (expectation value
in a state) satisfies $\u^2(p,q,t) = \u(p,q,t)$ which implies that
$\u(p,q,t)$ is the
characteristic function for some region; the quadratic profile ansatz
assumes that there is one connected region with boundary given by
$ (p- p_+(q,t) ) ( p - p_- (q,t)) = 0. $
As is well-known, in the ground state $| \psi_0 \ket$ the boundary is the
fermi surface itself: $ p^2 - q^2 - 2\mu=0,$ which corresponds to
$$ p_\pm^0 (q,t) = p_\pm^0 (q) = \pm \sqrt{q^2 + 2\mu}\;\vartheta (q^2 +
2\mu).
\eqn\thirtyfive $$
We shall use the notation
$$ p_\pm(q,t ) \equiv p_\pm^0 (q) + \eta_\pm(q,t)  \eqn\thirtysix$$
Let us compute  $\delt(u,v)$ in the state \thirtyfour\
{}~(ignoring the $O(\gst)$ terms for the moment). Using \twentysix,
\twentyseven\ ~and \thirtyfour\ ~--~ \thirtysix\ ~in \twelve, we get
$$\eqalign{
\delt(u,v)=& \int_{-\infty}^\infty dq\int^{\eta_+(q,t)}_0 {
dp \over |[2ue^t-(p+ p^0_+(q)+q)][2ve^{-t}- (p+p^0_+(q)-q)]|^{1/2} }
\cr
-& \int_{-\infty}^\infty dq\int^{\eta_-(q,t)}_0 {
dp \over |[2ue^t-(p+ p^0_-(q)+q)][2ve^{-t}- (p+p^0_-(q)-q)]|^{1/2} }
\cr }
\eqn\thirtyseven$$
Let us now make some further assumptions about the state $|\psi\ket$,
namely that the fluctuations $\eta_\pm(q,t)$ are non-zero only for $q <
-q_0 \equiv - \sqrt{-2\mu}$ and that in this region $|\eta_\pm(q,t)| <<
|p^0_\pm(q)|$. In other words, we are considering `small' fluctuations
near the left branch of the fermi surface in the classically allowed
region.
Under these assumptions we can expand the integrand in a Taylor series in
$p$ around $p=0$. The $p$-integrals now are easy to do, giving powers
of $\eta_\pm(q,t)$. The $q$-integrals that remain are now effectively
between $-\infty$ and $-q_0$, so one can use a
different integration variable $\tau$, the `time of flight', given by
$$ q = -q_0 \cosh \tau, \; p^0_\pm (q) = \pm q_0 \sinh \tau
\eqn\thirtyeight
$$
If we also define the  rescaled  functions $\etab_\pm(\tau, t) = | p_\pm^0
(q) | \eta_\pm (q,t) $, we ultimately get
$$\eqalign{
\delt&(u,v) ={1\over 2}
\int_0^\infty d\tau \Big[k_+(u,v|\tau,t) \etab_+(\tau,t) -
k_-(u,v | \tau, t)\etab_-(\tau,t)\Big] \cr
- {1\over 8}&
\int_0^\infty {d\tau\over p^0_+(q)}  (e^{-t}\del_u +e^t
\del_v) \Big[ k_+(u,v|\tau,t)
\etab_+^2(\tau,t)  -
k_-(u,v | \tau, t)\etab_-^2(\tau,t) \Big] + O(\etab_\pm^3)\cr}
\eqn\thirtynine$$
where
$$k_\pm(u,v| \tau,t) \equiv |(ue^t + {q_0\over 2}
e^{\mp \tau})(ve^{-t}- {q_0\over 2}e^{\pm
\tau })|^{-1/2}, \quad q_0 \equiv \sqrt{-2\mu}
\eqn\fourty$$

\noindent {\bf Relations between $\etab_\pm$ and $\delt(u,v)$:}

\nobreak
Equation \thirtynine ~is important in that it builds a correspondence
between the semiclassical quantities of the fermion theory and those of
the $\delt(u,v)$-theory. To understand it better, let us first choose a
different coordinatization for the $u,v$-space. Let us define\foot{
We use boldface letters so as to distiguish $\t$ from the time $t$ of the
fermion theory.}
$$ \x = {1\over 2} \ln |{uv\over \mu/2}|, \quad \t = {1\over 2}\ln |v/u|
\eqn\fourtyone$$
This is not a one-to-one map. Let us consider for the moment the quadrant
of the $u,v$-space where $u>0, v>0$. In that case we can write down the
inverse maps as follows:
$$ u = {q_0\over 2}e^{\x - \t},\quad v={q_0\over 2}e^{\x +\t}
\eqn\fourtytwo$$
Now recall that by definition of $\delt(u,v)$ ({\it cf.} the remark
about the $t$-independence of the right hand side of \twentytwo ), if
$\etab_\pm (\tau, t)$ satisfy their equations of motion, which can be
derived by tracing their definition back to $\u(p,q,t)$ and are
$$ (\del_t \mp \del_\tau)\etab_\pm= {1\over 2} \del_\tau \big(
{\etab_\pm^2 \over p_+^{0\,2} }
\big),
\eqn\fourtythree$$
then the right hand side of \thirtynine ~is actually $t$-independent. This
means that we can choose $t$ to be anything we like. It is most useful to
choose
$$ t = \t
\eqn\fourtyfour$$
on the right hand side of \thirtynine . This equation now reads,
to leading order, as
$$ \delt(\x,\t) = {1\over 2}
\int_0^\infty d\tau [\widetilde k_+(\x,\tau) \etab_+(\tau,
\t) -\widetilde k_-(\x,\tau) \etab_-(\tau, \t)] + O(\etab_\pm^2)
\eqn\fourtyfive $$
where
$$ \widetilde k_\pm (\x,\tau) \equiv {2\over q_0}
|(e^\x + e^{\mp\tau}) (
e^\x - e^{\pm\tau)}|^{-1/2}
\eqn\fourtysix $$
Note that for large $\x$,
$${q_0\over 2}\widetilde k_-(\x,\tau)\exp[\x] = |1 +
\exp(\tau - \x)|^{-1/2}+ O(e^{-\x})$$
The first term is  similar to a low-temperature
Fermi-Dirac distribution (the fact that the power is $-1/2$ instead of
$-1$ does not materially affect the arguments). In fact,
one can show that the
for very large $\x$ it behaves like $ \vartheta(\tau -\x)$ and
corrections to it are like increasing powers of $\del_\x$ on
the $\vartheta$-function. Similar expansions are also available for
$\widetilde
k_+(\x,\tau)$. The precise statements for these $\del_\x$-expansions are
the following:
$${\cal T}(\x,\t) = {1\over 2} \{{\cal D}_+\etab_+ - {\cal
D}_-\etab_- \} + O(e^{-\x} \etab_\pm) + O(\etab_\pm^2)
\eqn\fourtyseven$$
where
$$ {\cal T}(\x,\t) \equiv |uv|^{1/2} \delt(u,v) \eqn\fourtyeight$$
$$ {\cal D}_\pm \equiv I_\pm (\del_\x)+ I_\pm (1/2 - \del_\x)
\eqn\fourtynine$$
$$ I_\pm (\alpha) = \alpha^{-1} \;_2F_1({1\over 2}, \alpha; \alpha +1;
\mp 1) $$
where $\;_2F_1$ is the standard Hypergeometric function [\THIRTY].
Using its properties one can write down an expansion for ${\cal D}_\pm$ in
$\del_\x$. The expansion begins with $\del_\x^{-1}$. Defining
$$ \etab_\pm = \pi_\eta \pm \del_\tau \eta,
\eqn\fifty$$
where $\eta$ is the `tachyon' field that is associated with the standard
$c=1$ matrix model and $\pi_\eta$ is its conjugate momentum, we can write
down a derivative expansion for ${\cal T}(\x,\t)$:
$$ {\cal T}(\x,\t) = \eta(\x,\t) + O(\del_\x\eta, \pi_\eta).
\eqn\fiftyone$$
This identification of the `black hole tachyon' field with the standard
$c=1$ tachyon field in the asymptotic limit ($\x \to \infty$) is rather
remarkable. As a result of this, $n$-point functions of the opreator
${\widehat{\cal T}}(\x,\t)\equiv |uv|^{1/2} \ddelt(u,v)$
are the same as those of the standard $c=1$ tachyon at extreme low
energies.

Finally, relation \fourtysix\ ~can be inverted to give $\etab_\pm$
in terms of
${\cal T}$:
$$ \etab_\pm = ({\cal D}_\pm \del_\x)^{-1} \del_\pm {\cal T}(\x,\t)+
O(e^{-\x} {\cal T}) + O({\cal T}^2), \quad
\del_\pm = \del_\t \pm \del_\x
\eqn\fiftytwo$$
We will use this relation below to obtain a nonlinear differential
equation for ${\cal T}(\x,\t)$.

We wish to emphasize
that the above analysis can be repeated in other coordinates which are
valid all through the Kruskal diagram (for instance in the light cone
coordinates themselves). However, the formulae look more complicated.

\noindent {\bf Differential Equation:}

\nobreak
Let us now go back to the other consequences of \thirtynine.
Equation \thirty\  ~ensures that
$$ \Duv k_\pm (u,v | \tau, t) =0 $$
Using this, and applying the above differential operator to \thirtynine\ ~we
get
$$\eqalign {\Duv& \delt(u,v) \cr
={1\over 2}
&\del_u\del_v \int_0^\infty d\tau[k_+ \etab_+^2 + k_-\etab_-^2] +
O(\etab_\pm^3) \cr}
\eqn\fiftythree$$
Note that the linear term dropped out because of the special properties of
the kernel.
Using the $\x,\t$-coordinatization, \fiftythree ~can be written as
$$ D_{\x, \t} {\cal T}(\x, \t) = - {e^{-2\x}\over |\mu|} [
e^{ \x/2} \del_\x\{ e^{-\x/2}({\cal D}_+\etab_+^2 + {\cal D}_-\etab_-^2
)\}] + O(e^{-3\x}\etab_\pm^2) + O(\etab_\pm^3)
\eqn\fiftyfour$$
where
$$\eqalign{
D_{\x,\t} =& e^\x \Duv e^{-\x} \cr
=& (1+ e^{-2\x})(\del_\x^2 - \del_\t^2)+  e^{-2\x}(2\del_\x+1)\cr}
\eqn\fiftyfive $$
Finally, by using \fourtytwo\ ~to convert the $\etab_\pm$ back into ${\cal
T}$, we get the following closed differential equation  in ${\cal T}$ upto
quadratic order:
$$\eqalign{D_{\x, \t} {\cal T}(\x, \t) = - {e^{-2\x}\over |2\mu|} [
e^{ \x/2} \del_\x\{ e^{-\x/2}({\cal D}_+[({\cal
D}_+\del_\x)^{-1}\del_+{\cal T} ]^2 +&
{\cal D}_-[({\cal
D}_-\del_\x)^{-1}\del_-{\cal T} ]^2
)\}]\cr
\;&+ O({\cal T}^3) + O(e^{-3\x}{\cal T}^2)\cr}
\eqn\fiftysix$$
\bigskip

\noindent {\bf 5. \underbar{Exact Quantum Theory Does Not See Black Hole
Singuarity:}}

\nobreak
In this section we will analyze the nature of singularities that occur in
$\delt(u,v)$ at the position of the curvature
singularity of the black hole, $uv = \mu/2$. We will
see that these singularities occur as a result of making badly divergent
semiclassical expansions and they are not present in $
\delt(u,v) $ when calculated exactly.

Let us consider a state $| \psi \ket $ in the fermion theory which, like
in the previous section, represents fluctuations in the neighbourhood of
the left branch of the  fermi surface $p^2 - q^2 = 2\mu$ (generalizations
are obvious). In this region the fermion phase space can be coordinatized
by $(p,q) = (R \sinh \theta, -R\cosh \theta), R>0.$ As explained earlier,
the support of $\bra \psi | \uu(p,q,t) | \psi \ket $ in the limit $\gst
\to 0$ defines a region $\R(t)$ occupied by the fermi fluid at time $t$.
For the ground state $| \psi_0 \ket$  this region is given by $\R_0 = \{
\infty > R \ge R_0\equiv\sqrt{-2\mu}, \infty > \theta > -\infty \}$ (plus
its mirror image on the right half of the phase plane). For simplicity of
calculation, let us choose for the moment a state $| \psi \ket $ so that
the region $\R(t)$ has a particularly simple geometry. To be specific, we
choose that $\R(t=0)$ is obtained from $\R_0$ by adding a region $\delta
\R = \{ R_0 \ge R \ge R_1, \theta_1 \ge \theta \ge
\theta_2 \}$ and subtracting a
region $\widetilde{\delta\R} = \{ \widetilde R_1 \ge R \ge
R_0, \widetilde
\theta_1 \ge \theta \ge \widetilde \theta_2 \}$.
We will call $\delta \R$ the
``blip'' and $\widetilde {\delta \R}$ the ``antiblip''. The
state $| \psi \ket $ is created from the ground state $| \psi_0 \ket $ by
removing fermions from the region $\widetilde {\delta \R} \subset \R_0$
and placing them in the region $\delta \R$ just outside the filled fermi
sea. Fermion number conservation is achieved by choosing the areas of
$\delta \R$ and $\widetilde {\delta \R}$ to be the same, which is
equivalent to the condition that
$$ (\theta_1 - \theta_2) (R_1^2 - R_0^2) = (\widetilde{\theta_1} -
\widetilde {\theta_2}) (\widetilde R_0^2 - \widetilde R_1^2) $$
The region $\R(t)$ at non-zero times $t$ is simply obtained by shifting
the $\theta$-boundaries of both the blip and the antiblip by $t$. Using
this, one can easily write down the expression for $\delt(u,v)$ for this
state:
$$ \delt(u,v) = \delt_b(u,v) + \widetilde{\delt}_b (u,v) + O(\gst)
\eqn\fiftyseven
$$
where
$$ \delt_b (u,v) = {1\over 2}
\int^{R_0}_{R_1} RdR \int^{\theta_2}_{\theta_1} d\theta
| ( u + {R\over 2}e^{-\theta } ) (v - {R\over 2}
e^\theta )|^{-1/2}
\eqn\fiftyeight
$$
represents contribution of the `blip'.
The contribution of the `antiblip',
$\widetilde{\delt}_b(u,v),$ is a similar expression involving the
tilde variables. It is important to note that there are
$\gst$-corrections to
\fiftyseven\ ~since $\u(p,q,t)$ is actually a characteristic function plus
$O(\gst)$ terms.

A remark is in order about two seemingly different expansions that we are
making here. One is in $\gst$, and the other is in $|\delta
E/\mu|$. Ultimately, as it turns out, both are expansions in $\gst$. For
the moment, in \fourtyseven\ ~we have made an explicit $\gst$-expansion,
and no
$|\delta E/\mu|$ expansion yet. What we will show in the present
example is that
it is this latter expansion that is badly divergent and results in
increasingly singular behaviour at $uv = \mu/2$, and if one treats
expressions like \fourtyeight\ ~without a $|\delta E/\mu|$ expansion, then
$\delt(u,v)$ does not have any singularities at $uv = \mu/2$. More
generally, we will argue that singularities are invariably absent whenever
one performs the $|\delta E/\mu|$ resummation; the $\gst$-corrections
coming from corrections like those present in \fiftyseven\ ~do not affect the
conclusions {\it vis-a-vis} singularities, except in one extreme case
which we will discuss towards the end of this section.

Let us first analyze the singularities of \fiftyeight\ ~in the generic case
(treatment of $\widetilde
{\delt}_b
(u,v)$ is similar). Clearly singularities can arise only when the
expression inside the square root vanishes. It is also clear that a linear
zero inside the square root is not a singularity (recall that $\int_a^b
dx\, x^{-1/2}$ is not singular, for finite $a$ and $b$, even when the
origin is included in the range of integration),
we must have a quadratic zero. In other words,
both factors must vanish. This can happen only if $u<0, v>0$. Let us
choose the parametrization $u = -r e^{-\chi}/2, v= re^\chi/2$. We get
$$\delt_b(u,v) =
\int^{R_0}_{R_1} RdR \int^{\theta_2}_{\theta_1} d\theta | (R-r)^2 -
4Rr\sinh^2({\theta - \chi
\over 2})|^{-1/2}
\eqn\fiftynine$$
If we look at any one of the integrals separately, over $\theta$ or $R$,
we see a logarithmic singularity at $R=r,\theta= \chi$ provided this point
is included in the range of integration. Let us do the $\theta$ integral
first. If $r$ is outside the range $[R_1, R_0]$ we do not have any
singularities. If $r\in (R_1, R_0)$, and $\chi\in (\phi_1, \phi_2)$,
it is easy to see  that for $R \approx r$
the $\theta$ integral behaves as $\ln | R - r|$. Let us assume
that the range $[R_0, R_1]$ is small (compared to $R_0$, say, which means
that the blip consists of small energy fluctuations compared to the fermi
energy) so that $R\approx r$ throughout the range of integration (this is
only a simplifying assumption and the conclusions do not depend on it). We
get (for $\phi_2 > 1/2\ln(-v/u) > \phi_1$)
$$\eqalign{
\delt_b(u,v) \sim& \int_{R_1}^{R_0} RdR \ln |R -r| \cr
=& (R_0 - r)\ln |R_0 - r| - (R_1 - r) \ln |R_1 - r| - (R_1-R_0)\cr
\approx & (-\mu/2)^{-1/2}[ (uv - \mu/2)\ln| uv - \mu/2| - (uv - \mu/2 -
\Delta/2) \ln | uv - \mu/2 - \Delta/2| \cr}
\eqn\sixty$$
In the last line, we have put in the values $R_0 = \sqrt{-2\mu}, r =
\sqrt{ -4uv}$ and defined $R_1 \equiv \sqrt{2(-\mu - \Delta)}, \Delta> 0$
($\Delta $ thus measures the maximum energy fluctuation of the blip from
the fermi surface).  Since $R_0 \approx r \approx R_1$ we have used
$R_{0,1} - r \approx (R_{0,1}^2 - r^2)/(2R_0)$ and also $R_0 - R_1
\approx (R_0^2 - R_1^2)/(2R_0)$.

It is easy to see that the expression \sixty\ ~for
$ \delt_b(u,v)$ has no singularities\foot{
This statement is true all over $u,v$-space including the horizon.
}. However it is
also easy to see that it develops singularities as soon as one
attempts a semiclassical expansion in $\Delta/\mu$; to be precise one gets
$$\delt_b(u,v) \sim |\Delta/\mu| \ln| uv - \mu/2| + |\Delta/\mu|^2 (uv -
\mu/2)^{-1} + \cdots \eqn\sixtyone$$
where once again $1/2 \ln (-v/u) \in (\phi_1, \phi_2)$ (for $1/2 \ln
(-v/u)$ outside this range there are no singularities).

Thus, we see that the tachyon solution \fiftyeight\ ~develops a singularity at
$uv= \mu/2$ at the level of a $\Delta/\mu$ expansion, though the full
solution is regular.

What does the above example teach us for the general scenario?
In general, the tachyon solution would be given by expressions like
$$ \delt(u,v) ={1\over 2}\int RdR \int d\theta f(R,\theta) |
( u + {R\over 2} e^{-\theta}) ( v - {R\over 2}
e^\theta)|^{-1/2} \eqn\sixtytwo$$
where $f(R,\theta) \equiv \delta\u(R\sinh\theta, -R\cosh\theta)$ and  the
only thing that we have assumed is that the support of $\delta\u
\equiv \bra \psi |\uu | \psi \ket - \bra \psi_0| \uu | \psi_0 \ket $ is
confined to the region $p^2 < q^2$ (the generalization to the other
region is straightforward). Note that the expression \fiftyeight\ ~can be
recovered by setting the blip contribution to $\delta\u$, $\delta\u_b$, to
be
$$\delta\u_b = \vartheta [(R - R_0)(R_1 - R)] \vartheta [(\theta -
\theta_1)
(\theta_2 - \theta)] \eqn\sixtythree $$
where $\vartheta(x)= 1$ if $x>0$ and $0$ otherwise.
The integral in \sixtytwo\ ~consists of contributions from $R>0$ (left half
of
the phase plane) and from $R<0$. Let us look at the $R>0$ part first. Once
again the singularities can only come from $u<0, v>0$ region of the $u,v$
space. Using the same parametrization  as in \fourtynine\ ~we get
$$\delt(u,v) = \int RdR \int d\theta f(R,\theta) | (R-r)^2 -
4Rr\sinh^2({\theta - \chi\over 2})|^{-1/2}
\eqn\sixtyfour$$
The basic lesson of the previous example is that even though each
one-dimensional integral, taken separately over $R$ or $\theta$, has a
logarithmic singularity if the point $R=r, \theta=\chi$ is included in the
support of $f(R, \theta)$, the second integral smoothens out that
singularity. (In the previous example we did the $\theta$ integral first
to find logarithmic singularity and the $R$-integration smoothened that
out;
it could as easily have been done the other way around.) In fact the issue
is that of a {\bf two-dimensional} integration of the sort $\int dxdy\,
f(x,y)| x^2 - y^2 |^{-1/2}.$ For a smooth function $f(x,y)$ the possible
singularity at $x=y=0$ (a linear zero in the denominator) is washed away
by a stronger (quadratic) zero in the integration measure.
A singularity can be
sustained only if $f(x,y)$ has a pole or a stronger singularity at
$x=y=0$. However, in our case the quantum phase space density $\u(p,q,t)$
cannot have such singularities. The reason is that
the fermion field theory states
that we are concerned with are $W_\infty$-rotations of the ground state,
and therefore the corresponding $\u(p,q,t)$ is also a $W_\infty$-rotation
of the ground state density $\u_0(p,q)$. Since the latter is a smooth
distribution and a unitary rotation cannot induce  singularities,
we see that $\u(p,q,t)$ must be non-singular. The tachyon
field configurations
constructed by integrals such as \sixtyfour\ ~are therefore
non-singular too. Another way of
thinking about it is to use a two-step argument: the $\gst\to 0$ limit of
$\u(p,q,t)$ can be obtained from the $\gst \to 0$ limit of $\u_0(p,q)$
by a classical area-preserving diffeomporphism
(element of $w_\infty$) and is certainly nonsingular, being given by the
characteristic function of a region equal to the $w_\infty$-transformed
Fermi sea. This implies that the tachyon field constructed from it is
already non-singular. Incorporating  $\gst$-effects further smoothens
out these distributions and hence cannot introduce singularities in the
tachyon field.

Let us now discuss the exceptional case mentioned briefly at the begining
of this section.  This situation happens in the extreme case when the blip is
infinitely thin so that it must spread out over an entric hyperbola in
$\theta$ (to conserve fermion number).  In this case the $R$-integration
in \fiftyeight\ ~drops out and so there is no $|\delta E/\mu|$ expansion.  The
$\gst$ corrections, therefore, come entirely from \fiftyseven .  It is
these, when summed to all orders, that are responsible for washing out the
black hole singularity in this case.

To illustrate this let us consider the example of the quantity
$\partial_\mu \langle \psi_0|\hat T(u,v)|\psi_0\rangle \equiv \partial_\mu
T_0(u,v)$.  Physically this quantity represents the change in the vacuum
expectation value of the tachyon field $\hat T(u,v)$ as the fermi level is
shifted a little bit.  Since $\langle \psi_0|\uu(p,q,t)|\psi_0\rangle =
\theta\left(\mu - {p^2 - q^2 \over 2}\right)$ in the $\gst \rightarrow
0$ limit, we see that in this limit
$$
\partial_\mu T_0 (u,v) \rightarrow {1\over2} \int^{+\infty}_{-\infty} RdR
\int^{+\infty}_{-\infty} d\theta \Big|\left(u + {R\over2}
e^{-\theta}\right) \left(v - {R\over2} e^\theta\right)\Big|^{-{1\over2}}
\delta \left(\mu + {R^2 \over 2}\right)
$$
Because of the $\delta$-function in the integrand the $R$-integration
drops out and one can check that the $\theta$-integral diverges as $\ell
n\left(uv - {\mu \over 2}\right)$ in the limit $uv \rightarrow \mu/2$.  The
full quantum answer for $\partial_\mu T_0(u,v)$ is, however, nonsingular
[\TWENTYTWO].  In fact,
$$
\eqalign{
\partial_\mu T_0(u,v) = \sqrt{\pi\over2} Im \int^\infty_0 d\lambda &
{e^{i\mu \lambda - 2i u v \tanh {\lambda\over2}} \over \sqrt{\sinh
\lambda}} \Bigg[e^{-{\pi\over4}} H_0^{(1)} \left(2|uv|\tanh
{\lambda\over2}\right) \cr &
+ e^{i{\pi\over4}} H^{(2)}_0 \left(2|uv|\tanh {\lambda \over 2}\right)\Bigg]}
$$
where $H^{(1)}_0$ and $H^{(2)}_0$ are standard combinations of Bessel
functions [\THIRTYONE].  The $\gst$ expansion is obtained by the
rescalings $uv
\rightarrow uv/\kappa$ and $\mu \rightarrow \mu/\kappa$ and expanding in
powers of
$\kappa^2$.  Doing so and retaining only the lowest order term, we recover the
earlier logarithmic singularity at $uv = \mu/2$.  It can, however, be
easily verified that the exact expression is nonsingular at $uv = {\mu
\over 2}$.

We conclude, therefore, that the exact quantum theory does not permit any
singlarities in the tachyon field configuration.
\bigskip

\noindent {\bf 6. \underbar{\bf Some Novel Features of the Quantum Field
Theory of $\T(u,v)$}}

\nobreak
So far we have discussed the semiclassical physics of $\delt(u,v)$
including its low energy differential equation  and in
the last section we have seen how to compute the expectation values of
$\ddelt(u,v)$ beyond the semiclassical expansion using the fermion theory.
In this section we discuss in more detail some novel features of
the two-dimensional quantum field theory of $\ddelt(u,v)$ defined by the
underlying fermion theory.

Let us first discuss how $\ddelt(u,v)$, which is
{\it a priori} defined in terms of an on-shell three-dimensional field
$\uu(p,q,t)$, defines a Heisenberg operator in a {\bf two-dimensional}
field theory. Basically we use the fact that the right hand side of
\twentytwo\ ~is actually independent of $t$ to put $t$ equal to the
`time' of the $u,v$-space. To fix ideas, let us consider the
coordinatization \fourtyone\ - \fourtytwo\ of the $u,v$-space where $\x,\t$
correspond to space and time. This of course limits the present
discussion to the
quadrant $\{u>0, v>0\}$, but a similar discussion holds
for more global choices
of `time' coordinates also. Now, in these coordinates,
writing $\ddelt(u(\x,\t), v(\x,\t))$ as $\ddelt(\x,\t)$ by an abuse of
notation, we have
$$\eqalign{
\ddelt(\x,\t)& \; \cr
= \int dudv&\;
| (e^\x-ue^\t)(e^\x-ve^{-\t})|^{-1/2}\delta \uu(ue^t + ve^{-t}, ue^t -
ve^{-t}, t) \cr
= \int dudv&\;
| (e^\x-u)(e^\x-v)|^{-1/2}\delta \uu(ue^{t-\t}+ ve^{\t-t},
ue^{t-\t}-ve^{\t-t},
t) \cr}
\eqn\sixtyfive$$
Here
$$ \delta \uu(p,q,t) \equiv \uu(p,q,t) - \bra \psi_0 | \uu(p,q,t) | \psi_0
\ket $$
Since
these expressions are actually independent of $t$, we can choose the
`gauge'
$$ t = \t
\eqn\sixtysix$$
which gives
$$\ddelt(\x,\t) = \int dudv\; |(e^\x - u)(e^\x -v)|^{-1/2}
\delta \uu(u+v,u-v, \t) \eqn\sixtyseven$$
Note that the fields on both sides of \sixtyseven\ ~are evaluated at the
{\bf same}
time $\t$. More explcitly we see that
$$\ddelt(\x,\t) = e^{iH\t} \ddelt(\x,0) e^{-iH\t} \eqn\sixtyeight$$
where
$$\ddelt(\x,0) = \int dudv\; |(e^\x - u)(e^\x -v)|^{-1/2}
\delta \uu(u+v,u-v, 0)
\eqn\sixtynine$$
The hamiltonian is the same as in \fifteen.  Equations \sixtyeight\ ~and
\sixtynine\
{}~tell us that
$\ddelt(\x,\t)$ is a Heisenberg operator in a two-dimensional field
theory. Equation \sixtyseven\ ~allows us to write down time-ordered products of
$\ddelt(\x,\t)$'s in terms of time-ordered products of the $\delta
\uu(p,q,\t)$'s:
thus
$$\eqalign{
\; &  {\cal T}(\ddelt(\x_1,\t_1)\cdots \ddelt(\x_n,\t_n)) \cr
=&\int du_1dv_1\cdots du_ndv_n |(e^{\x_1}-u_1)(e^{\x_1}-v_1)|^{-1/2}
\cdots
|(e^{\x_n}- u_n)(e^{\x_n} - v_n)|^{-1/2}\times \cr
\qquad &{\cal T}(
\delta \uu(u_1+v_1, u_1-v_1,\t_1)\cdots \delta \uu(u_n + v_n, u_n - v_n,
\t_n))
\cr}
\eqn\seventy$$
In this way expectation values of time-ordered products of
$\T(u,v)$ can be computed from the fermion theory. As we have stressed
earlier, in principle these contain answers  to all dynamical
questions in the theory. However, these correlation functions are not
related  to usual particle-scattering amplitudes in the standard fashion,
parimarily because:

$\bullet$ $[\ddelt(\x_1,\t), \ddelt(\x_2, \t)] \ne 0$, {\it i.e.},
the field $\ddelt(\x, \t)$ does not commute with itself at equal times.

In fact the non-trivial commutation relation is a direct consequence of
the $W_{\infty}$ algebra.  As we should expect, the field $\T(u,v)$ bears
a close resemblance to the spin operator in a magnetic field. In both
cases the symplectic structures (ETCR in the quantum theory) are
non-trivial. We know that in case of the spin, dynamical questions are
better formulated in terms of coherent states $| {\bf n}\ket$ satisfying
$\bra {\bf n} |\widehat{\bf S} | {\bf n} \ket = {\bf n}$. Questions such
as how $| {\bf n} \ket$ evolves in time are equivalent to calculating the
dynamical trajectory $\bra \widehat {\bf S(t)} \ket $. In the present case
$\delt(u,v) \equiv \bra \ddelt(u,v) \ket$
plays a role exactly similar to this object.

$\bullet$ Two-point function $\ne $ ``Propagator''.

We have seen in Sec. 4 that $\delt(u,v)$, or equivalently ${\cal
T}(\x,\t)$, satisfies the black hole differential equation \fiftysix .
In an ordinary scalar field theory such a thing would imply
$$ D_{\x,\t} G_2(\x,\t| \x',\t') = (1+ e^{-2\x})
\delta(\x-\x') \delta (\t-\t') + O(\gst) \eqn\seventyone$$
where
$$
G_2(\x,\t|\x',\t') \equiv \bra \psi_0 | T(
{\widehat{\cal T}}(\x,\t) {\widehat{\cal T}} (\x',\t') | \psi_0 \ket
\eqn\seventytwo$$
The prefactor $(1 + \exp(-2\x))$ is equal to  $1/\sqrt
{\hbox{det}\, g}$ in $(\x,\t)$ coordinates, required to make the
$\delta$-function covariant. The operator
${\widehat{\cal T}}$ is defined as
$$\widehat{\cal T}(\x, \t) \equiv | uv |^{1/2} \ddelt(u,v).
\eqn\seventythree$$
The symbol
$T$ in \seventytwo\ ~denotes time-ordering in the time $\t$.

In our case, \seventyone\ ~is not true because of the non-standard
commutation relations of the $\widehat{ \cal T}$ field. It is easy to
derive that
$$
D_{\x, \t}G_2(\x, \t| \x', \t')  = - (1 + e^{-2\x}) \big( \delta (\t -
\t') [ \del_t{\widehat {\cal T}} (\x, \t), {\widehat {\cal T}} (\x', \t)]
+ \del_\t \delta(\t - \t') [ {\widehat {\cal T}} (\x, \t'), {\widehat
{\cal T}} (\x', \t') ] \big)
\eqn\seventyfour$$
The commutation relations of the
${\widehat {\cal T}}$ fields can be derived from the definition
\seventythree\
{}~and the $\uu(p,q,t)$ commutation relations.
Neglecting corrections of order $e^{-\x}$ we have
\def\ttt{{\widehat {\cal T}}}
\def\ddd{{\cal D}}
$$\eqalign{
[\ttt (\x, \t), \del_\t \ttt(\x' ,\t) ]
&\propto [\ddd_+' \ddd_+ + \ddd_-'
\ddd_-] \del_\x^2 \delta (\x - \x') \cr
[\ttt (\x, \t), \ttt(\x' ,\t) ] &\propto [\ddd_+' \ddd_+ - \ddd_-'
\ddd_-] \del_\x \delta (\x - \x') \cr}
\eqn\seventyfive $$
where $\ddd_\pm$ are as defined in \fourtynine\ ~(the primes refer to
$\x'$).
In the limit of extremely large $\x$,  the operators $\ddd_\pm$ go as
$(\del_\x)^{-1}$ and we recover canonical communication relations. This is
essentially because in this limit the field $\ttt(\x,\t)$ becomes the same
as the `tachyon' $\eta$ of the standard $c=1$ matrix model (see \fiftyone\ ).

The non-identification of the two-point function with the propagator is
basically related to the fact that $\ddelt(u,v)$ or $\ttt(\x,\t)$ cannot
create particle states because they do not commute at equal times.

$\bullet$ How does one address the issue of propagation then?

As we have stressed, $\T(\x, \t)$ can be regarded as a Heisenberg
operator in a two-dimensional field theory.
If $H$ is a functional of $\T(\x,0)$ and $\del_\t \T(\x,0)$, then
given $\bra \psi | \T(\x,0) | \psi \ket$ and
$\bra \psi | \del_\t\T(\x,0) | \psi \ket$, one can in
principle determine $\bra \psi | \T(\x, \t) | \psi \ket $. In general it
is a
difficult question whether $H$ is a functional of only $\T(\x,0),
\del_\t \T(\x,0)$. However,
even if it is not clear how much initial data is
required to get a unique dynamical trajetory (unique answer for, let's
say, $\bra \psi | \T(\x, \t) | \psi \ket $), the fermionic
construction {\bf does} list {\bf all} dynamical trajectories.
Also, we do know from the analysis of small fluctuations
(for instance using the language of $\etab_\pm$) that data worth two real
functions (e.g. $\etab_\pm(\tau, t=0)$) are enough to determine the future
evolution for the fermionic state, except perhaps for some discrete data
corresponding to `discrete states'.
Incidentally, it is also clear from the
one-to-one correspondence between $\etab_\pm(\tau)$ and $\T(\x,0),
\del_\t \T(\x,0)$ that we can choose {\bf any} kind of initial data
on any given spacelike (or lightlike) surface $\t = \t_0$ (this
list includes the white hole horizon) by simply choosing the appropriate
fermionic state, or equivalently the appropriate values for
$\etab_\pm(\tau,\t_0)$.
\bigskip

\noindent  {\bf 7. \underbar{\bf Analytically continued Fermion Theory and
The Euclidian Black Hole}}

\nobreak
In this section we
would like to study the implication for the black hole
tachyon theory of the
analytic continuation of the Fermion field theory discussed earlier in
[\SIXTEEN], namely\foot
{This is equivalent to the analytic continuation
[\THIRTYTWO]
of the harmonic oscillator frequency $w \to iw$ in the
fermion potential.}
$$ t \to it, \; p \to -ip \eqn\seventysix$$
In this analytic continuation,
the `hyperbolic transform' becomes an `elliptic transform',
$$ \pphi(p,q,t) = \int { dp' dq' \over [(p - p')^2 + (q - q')^2 ]^{1/2} }
\uu (p,q,t), \eqn \seventyseven$$
Then, by the analytically continued equation of motion,
$$  (\del_t + p \del_q - q \del_p) \uu(p,q,t) = 0 \eqn\seventyeight$$
we can show that $\pphi (p,q,t)$ is of the form
$$ \pphi (p,q,t) = \T(u, \bar u)       \eqn\seventynine$$
where
$$ u= { q - ip \over 2} \exp( -it),\; \bar u = {q + ip \over 2} \exp (it)
\eqn\eighty$$
As in Sec. 3,  one can show that in the low energy
approximation $\delt(u, \bar u)$ satisfies the equation of motion of a
massless field in the Euclidian black hole:
$$ [\barDuv ] \delt(u, \bar u) = 0 + O((\delt)^2)  \eqn\eightyone$$
This corresponds to a dilaton-metric background that describes the
Euclidian black hole (the `cigar').

It is remarkable that the analytic continuation
\seventysix\ ~of the matrix model
defines the usual analytic continuation of the black hole physics!
This makes it tempting to believe that the thermal Green's functions of
the latter
may have a direct significance in terms of the matrix model in the
forbidden region.
\bigskip

\noindent {\bf 8. \underbar{Concluding Remarks}}

\nobreak
The reinterpretation of the $c=1$ matrix model as a model of black holes
presented and discussed in detail in the preceeding pages has raised many
conceptual issues that need deeper understanding.

Firstly there is the question of the $S$-matrix. The definition of the
$S$-matrix requires the specification of the `in' and `out' states. These
states can be inferred by analyzing the two-point correlation function of
a complete commuting (equal time) set of field operators.
In the present case a natural set is
the density operator $\psi^\dagger (x, t)\psi (x,t) \equiv \del_x
\varphi$, because perturbatively we know that $\varphi$ creates `in' and
`out' massless particle states. Once this is done we can evolve the `in'
states to the `out' states in the standard fashion and define the
$S$-matrix.  For the $c=1$ theory some $S$-matrix elements have been
previously calculated
[\THIRTYTHREE,\TWELVE,\TEN,\THIRTYFOUR].
The black hole interpretation of this theory, in particular
the non-linear differential equation \fiftysix, seems to strongly suggest
that the kernel that propagates `in' states to the `out' states has a
representation in the $T(u,v)$ theory. It would be very interesting to see
this explicitly.

The other aspect of the present work that may at first appear puzzling is
that the $c=1$ matrix model seems to have a two-dimensional space-time
interpretation in terms of two different metrics.  As we have emphasized in the
Introduction, to appreciate this point one must remember that the
interpretation
of the $c=1$ matrix model in terms of a two-dimensional space-time is not
an a priori given thing and, therefore, an interpretation in terms of a
unique space-time metric is not guaranteed.  Indeed, by actually
constructing the black hole interpretation we have precisely realized this
possibility in the present work.  We believe that this phenomenon is
deeply connected with the disappearance of the classical black hole
singularity in the present model.  It would, threfore, be very interesting
to see if it can be generalized to more realistic number of dimensions.

One of the limitations of the $c=1$ matrix model as a model of black holes
is that this black hole is eternal and one cannot envisage any process
({\it e.g} formation and evaporation) that involves changing the mass of
the black hole. The simple reason  for this is that the mass of the black
hole is equal to the fermi level which cannot be changed if the number of
fermions is held fixed. It is an interesting question whether there is a
way of circumventing this difficulty within the context of string theory.
It does, however, seem possible to study the other phenomenon in the
present model, namely, excitation and deexcitation of the black hole.  One
could imagine throwing in matter at the black hole and studying the
resulting back reaction.  One way of studying this would be to look at
tachyon fluctuations around excited states of the fermion theory.  The
reason for this is that since the eternal black hole corresponds to the
fermi ground state, throwing in matter at the black hole must correspond
to going to an excited state.  We may then repeat the calculations of Sec. 4
with the difference that $p^0_\pm (q,t)$ are no longer given by
\thirtyfive.  Instead, these now satisfy the Euler equation
$$
\partial_t p^0_\pm = p^0_\pm \partial_q p^0_\pm - q,
$$
on which one may impose the boundary condition that in the infinite past
$p^0_\pm (q,t)$ reduce to the $p^0_\pm (q)$ of equation \thirtyfive.  One
could go ahead and derive an equation analogous to \thirtynine\ for
fluctuations around this background.  At this stage it is difficult to proceed
any further without making approximations since the analogues of the
kernals $k_\pm$ in this case are
not solutions of the black hole differential equation.  However, in the
approximation that $p^0_\pm (q,t)$ differ from the fermi ground state by
only a `small' amount at all times, one may use double perturbation theory,
further expanding the kernals around the fermi ground state.  The
differential equation one gets for the fluctuations in this way is
precisely the one that could be obtained directly by expanding \fiftysix\
around a `small' nontrivial background ${\cal T}_0$.  Clearly, every term
on the
right hand side of \fiftysix\ contributes to the piece linear in fluctuations
around ${\cal T}_0$ and, therefore, the differential operator acting on the
fluctuation on the left hand side is modified from the one appearing in
\fiftysix.  ~~One effect of this is to modify the two-dimensional metric,
in particular, to a time-dependent one.  The other effect is to introduce
higher than quadratic derivatives, which might find an explanation in
terms of backgrounds of higher tensor fields (discrete states) in addition
to the metric.  It would clearly be worthwhile to explore these issues in
greater detail.

\bigskip

\noindent \underbar{Acknowledgements:}  It is a pleasure to thank the
organizers of the Spring Workshop for the invitation and for hospitality
during my stay at ICTP.  The work described in these notes was done in
collaboration with G. Mandal and S.R. Wadia.

\endpage

\refout

\end